\documentclass{emulateapj} 
\usepackage{apjfonts}

\newcommand\beq{\begin{equation}}
\newcommand\eeq{\end{equation}}
\def\rmmat#1{{\hbox{\rm #1}}}

\newcommand{\etal}{et al.}

\def\rosat{{\it ROSAT}}
\newcommand\suzaku{{\it Suzaku}}
\newcommand\chandra{{\it Chandra}}

\newcommand\xmm{{\it XMM-Newton}}

\def\psr{\hbox{PSR~J0821$-$4300}}
\def\pupa{\hbox{Puppis~A}}
\def\snr{\hbox{Puppis~A}}

 \def\pkssrc{{\rm PKS~1209$-$51/52}}

\def\bestpsi{86^{\circ}}
\def\bestxi{6^{\circ}}

\begin{document}
 
\title{Modeling the Surface X-ray Emission and Viewing Geometry of \psr\ in Puppis A}

\author{E.~V.~Gotthelf\altaffilmark{1}, R.~Perna\altaffilmark{2} and J.~P.~Halpern\altaffilmark{1}} 
\altaffiltext{1}{Columbia Astrophysics Laboratory, Columbia University, New York, NY 10027}
\altaffiltext{2}
{JILA and Department of Astrophysical and Planetary Sciences, University of Colorado, Boulder, CO, 80309}

\begin{abstract}
We present a model for the unusual X-ray pulse profiles of \psr, 
the compact central object in supernova remnant \pupa. 
We show that a pair of thermal, antipodal hot-spots on the neutron
star surface is able to fully account for the pulsar's double
blackbody spectrum and energy-dependent pulse profile, including the
observed $180^{\circ}$ phase reversal at $\approx 1.2$~keV. By
comparing the observed pulse modulation and phase to the model
predictions, we strongly constrain the hot-spot pole ($\xi$) and the
line-of-sight ($\psi$) angles with respect to the spin axis. For a
nominal radius of $R =12$~km and distance $D = 2.2$~kpc, we find
$(\xi,\psi) = (\bestpsi,\bestxi)$, with $1\sigma$ error ellipse of
($2^{\circ},1^{\circ}$); this solution is degenerate in the two
angles.  The best-fit spectral model for this geometry requires that
the temperatures of the two emission spots differ by a factor of 2 and
their areas by a factor of $\sim 20$.  Including a cosine-beamed
pattern for the emitted intensity modifies the result, decreasing the
angles to ($84^{\circ},3^{\circ}$); however this model is not
statistically distinguishable from the isotropic emission case.
We also present a new upper limit on the period derivative of
$\dot P < 3.5 \times 10^{-16}$ ($2\sigma$), which limits the
global dipole magnetic field to $B_s < 2.0 \times 10^{11}$~G,
confirming \psr\ as an ``anti-magnetar.''  We discuss the
results in the context of observations and theories of nonuniform
surface temperature on isolated NSs of both weak and strong magnetic
field. To explain the nonuniform temperature of \psr\ may require a crustal 
field that is much stronger than the external, global dipole field.
\end{abstract}


\section{Introduction}

The young X-ray pulsar \psr, associated with the \pupa\ supernova
remnant (SNR), is one of three pulsars in SNRs
that are spinning down nearly imperceptibly \citep[][Paper I]{got09}.
The age of \psr\ is $3.7$~kyr based on the
proper motion of oxygen knots in \snr\ \citep{win88}, and
its distance is $2.2$~kpc from \ion{H}{1} absorption features
to the SNR \citep{rey95}.
In the context of the magnetic dipole model, these pulsars
are a new physical manifestation of neutron stars (NSs),
``anti-magnetars'' born with weak
magnetic fields possibly related to slow natal spin \citep{got08}.
They were drawn from the previously defined
class of central compact objects (CCOs) in SNRs,
which are characterized by their steady, predominantly thermal X-ray emission,
lack of optical or radio counterparts, and absence of pulsar wind
nebulae (see reviews by
\citealt{pav04} and \citealt{del08}). Currently, 7--10
objects are known or proposed to be CCOs 
\citep[for a list, see][]{hal10},
and are therefore candidates for anti-magnetars.

The \xmm\ discovery observations of \psr\ revealed an
abrupt $180^{\circ}$ phase reversal of its quasi-sinusoidal
pulse profile at an energy of around $1.2$~keV (Paper~I).
The X-ray spectrum of \psr\ was
fitted with a two-temperature blackbody model,
both temperatures being seen at all rotation phases,
while the cross-over energy of the spectral components
coincides with the energy where the pulse reverses phase.
These detailed properties afford the opportunity to construct
a highly constrained model of the NS surface emission geometry.

In this Paper, we present a quantitative verification of the
geometrical model for \psr\ proposed in Paper~I.  By reproducing the
detailed pulse profile behavior, we are able to specify the surface
emission areas and viewing geometry to within $< 2^{\circ}$.  Our
treatment includes general relativistic effects of light deflection
and gravitational redshift, and examines the effects of local
anisotropy (beaming) in the emitted radiation.  We describe the
antipodal hot-spot model in Section~2, compare the energy-dependent
modulation to the data for the range of allowed geometries in
Section~3, and explore if it is possible to limit the neutron star
radius in conjunction with the estimated distance.  In Section~4 we
discuss some implications of the model results, and in Section~5
compare with pulsars of other types.

\section{The Emission model}

\begin{figure}[t]
\plotone{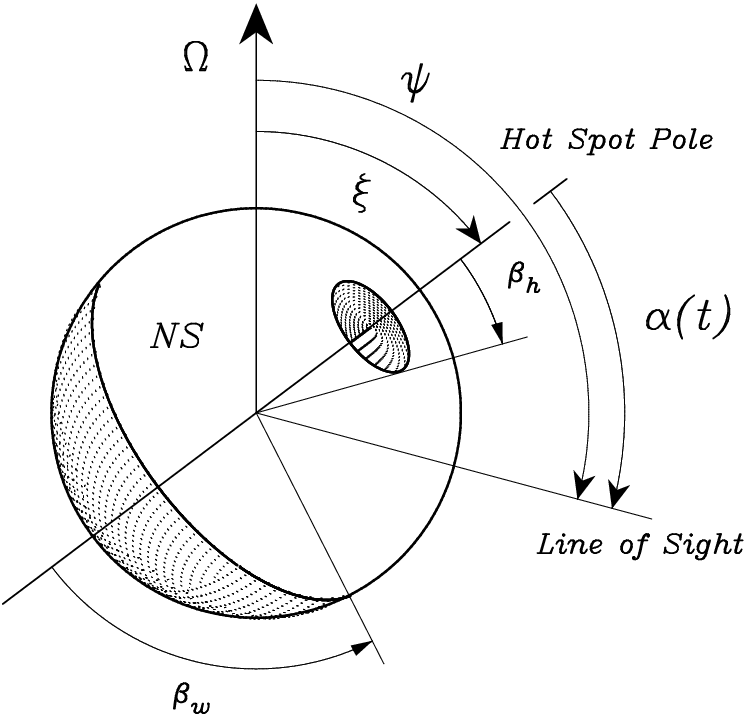}
\caption{Emission geometry on the surface of the NS for
the model presented herein: a hot spot of temperature $T_h$ and
angular size $\beta_h$ and an antipodal spot of temperature
$T_w$ and angular size $\beta_w$. As the NS
rotates with angular velocity $\Omega$, the angle $\alpha(t)$ is a
function of the phase angle $\gamma(t)=\Omega t$, the angle
$\xi$ between spin axis and hot spot axis, and the angle $\psi$
between spin axis and line-of-sight.}
\label{fig:NS}
\end{figure}

Our method of modeling the emission from spots on the surface of
a NS follows the derivation given by \citet{pec83}, 
with some generalizations introduced by \citet{per08}.
The radiation comes from a hot spot of blackbody temperature $T_h$ and
angular radius $\beta_h$, and an antipodal warm spot of
lower temperature $T_w$ and angular radius $\beta_w$.
The remainder of the surface is assumed to be at a uniform
temperature $T_{\rm NS} < T_w$.
The geometry is indicated in Figure~\ref{fig:NS}.

We use $\gamma(t)$ to indicate the phase of rotation
instead of the common notation $\phi(t)$.
Phase $\gamma=0$ corresponds to the closest
approach of the hot spot to the observer, while the phase
of rotation is related to the angular rotation
rate of the star $\Omega$ through $\gamma(t)=\Omega t$.
We indicate with $\alpha_h(t)$ the angle that the hot-spot axis
makes with the line-of-sight. $\alpha_h(t)$ is a function of the angle $\xi$
between the hot-spot axis and the rotation axis and the angle $\psi$
between the line-of-sight and the rotation axis, by means of the relation
\beq
\alpha_h(t)=\arccos[\cos\psi\cos\xi+\sin\psi\sin\xi\cos\gamma(t)]\;.
\label{eq:alpha}
\eeq
For each set of angles $\xi$ and $\psi$, the angle $\alpha_w$ that
the axis of the opposing warm spot makes with respect to the line-of-sight
is simply $\alpha_w(t)=\pi-\alpha_h(t)$.

The spherical coordinate system $(\theta,\phi)$ is 
defined with respect to the line-of-sight as the $z$-axis.
Due to general relativistic effects,  
a photon emitted at a colatitude $\theta$ reaches the observer
only if emitted at an angle $\delta$ with respect to the perpendicular to the NS surface. 
The two angles are related  by the
ray-tracing function\footnote{To improve the computational efficiency
of this equation we use the approximation presented in
\citet{bel02}.} \citep{pec83,pag95a}
\beq
\theta(\delta)=\int_0^{R_s/2R}x\;du\left/\sqrt{\left(1-\frac{R_s}{R}\right)
\left(\frac{R_s}{2R}\right)^2-(1-2u)u^2 x^2}\right.\;,
\label{eq:teta}
\eeq 
having defined $x\equiv\sin\,\delta$. Here, $R/R_s$ is the ratio of the
NS radius
to Schwarzschild radius, $R_s=2GM/c^2$ (we will assume $M=1.4\,M_\odot$). 

The hot spot is bounded by
the conditions: 

\beq
\theta\le\beta_h\;\;\;\;\;\;\;\;\;\;\;\; \rmmat{if}\;\;\; \alpha_h=0\;
\label{eq:con1}
\eeq
and
\beq
   \left\{
  \begin{array}{ll}
    \alpha_h-\beta_h\le\theta\le\alpha_h+\beta_h \\
      2\pi-\phi_p^h\le\phi\le\phi_p^h \;\; \;\;\;\;\rmmat{if}
        \;\;\;\alpha_h\ne 0\;\;\;\rmmat{and} \;\;\;\beta_h\le\alpha_h,\\
  \end{array}\right.\;
\label{eq:con2}
\eeq
where
\beq
\phi_p^h=\arccos\left[\frac{\cos\beta_h-\cos\alpha_h\cos\theta}{\sin\alpha_h\sin\theta}\right]\;.
\label{eq:phip}
\eeq
On the other hand, it is identified through the condition 
\beq
\theta\le\theta^h_*(\alpha_h,\beta_h,\phi)\;\;\;\;\;\ 
\rmmat{if}\;\;\; \alpha_h\ne 0\;\;\;\rmmat{and}\;\;\;\beta_h > \alpha_h\;,
\label{eq:con3}
\eeq
where the outer boundary $\theta^h_*(\alpha_h,\beta_h,\phi)$ of the spot is computed by numerical
solution of the equation 
\beq
\cos\beta_h = \sin\theta_*^h\sin\alpha_h\cos\phi + \cos\theta_*^h\cos\alpha_h\;.
\label{eq:t*}
\eeq
The antipodal warm spot is described on the surface of the star through
the same conditions, but with the substitutions  $\beta_h\rightarrow\beta_w$
and $\alpha_h\rightarrow\alpha_w$.

We assume that the emission from the hot and warm spots is blackbody,
of uniform temperatures $T_h$ and $T_w$, respectively.  The spectral
function is then given by $n(E,T)=1/[\exp(E/kT)-1]$, where the
temperature $T(\theta,\phi)$ is equal to $T_h$ or $T_w$ if $\theta$
and $\phi$ are inside either of the spots, respectively, and it is
equal to zero outside. Given that the presence of NS atmospheres and
their elemental composition are yet to be firmly established, here we
first model isotropic radiation.  Then, we explore how the obtained
constraints depend on the assumption of forward beaming of the
radiation, as found in magnetized, light element atmosphere models
\citep[e.g.,][]{pav94}, by approximating beaming\footnote{This
approximation is based on the assumption that the magnetic field is
normal to the surface in the heated regions; this is plausible since
heat transport along the B field lines is enhanced in the outer
envelope, at least for $B\ga 10^{10}$~G. See
\S4 for further discussion of this issue.} as intensity
$I(\delta) \propto \cos\,\delta^n$.  The observed spectrum as a
function of phase angle $\gamma$ is obtained by the standard method of
integrating the local emission over the observable surface of the
star, accounting for the gravitational redshift of the radiation
following \citet{pag95a}:

\begin{eqnarray}
F(E_\infty,\gamma) =\frac{2 \pi}{c\,h^3}\frac{R_\infty^2}{D^2}\;E_\infty^2
e^{-N_{\rm H}\sigma(E_\infty)} \int_0^1 2xdx\nonumber \\ 
 \times \int_0^{2\pi} \frac{d\phi}{2\pi}\; 
I_0(\theta,\phi) \;n[E_\infty e^{-\Lambda_s};T(\theta,\phi)]\;
\label{eq:flux}
\end{eqnarray} 
in units of photons cm$^{-2}$ s$^{-1}$ keV$^{-1}$.  In equation~(\ref{eq:flux}),
the NS radius and photon energy as observed at infinity are given by
$R_\infty= Re^{-\Lambda_s}$ and $E_{\infty}= E e^{\Lambda_s}$, where
$E$ is the energy emitted at $R$,
and ${\Lambda_s}$ is defined as
\beq
e^{\Lambda_s}\equiv\sqrt{1-{\frac{R_s}{R}}}.  
\eeq 
The phase-averaged flux is computed as $F_{\rm avg}(E_\infty)=
1/2\pi\int_0^{2\pi}d\gamma F(E_\infty,\gamma)$. 

In addition to the basic two-temperature antipodal spot model
described above, the spectrum of \psr\ requires an additional narrow
line-like component around 0.77 keV, possibly an electron cyclotron
feature in emission (Paper~I). Furthermore, as shown in Paper~I, this
spectral feature is associated exclusively with the larger spot, of
temperature $T_w$.  In the current study, we include the best fitted
Gaussian line as an additive component to our basic model. With no
other information about its spatial distribution, this emission is
assumed to be spread uniformly over the surface of the warm spot
only. This line emission is shown to account for a notable increase in
the observed modulation below 1~keV, as described in Section~3.

\section{Modeling the Energy-Dependent Modulation}

The surface emission geometry of \psr\ is highly constrained by its
unique energy dependent pulse profile.  As shown in Paper~I, the
quasi-sinusoidal signal has a background subtracted pulsed fraction of
$\approx 11\%$ in the energy band $0.5-4.5$~keV, with an abrupt
$180^{\circ}$ change in phase at $1.2$~keV, around which the
modulation evidently cancels out.  This behavior is indicative of a
geometry having the symmetry of Figure~\ref{fig:NS}, namely, a pair of
antipodal spots of different temperatures.  Our goal is to match the
observed pulse profile (modulation and phase) in three
interesting energy bands, $0.5-1.0$, $1.0-1.5$, and $1.5-4.5$~keV, using
the antipodal model, by exploring the range of all possible viewing
and hot-spot geometry pairs ($\xi,\psi$; see Figure~\ref{fig:NS}),
and fitting for the correct one.  

We summarize our proceedure as follows. We started by fitting the
X-ray spectrum of Paper~I using the antipodal model to compute the
temperatures and spot-sizes corresponding to all spot and viewing
angles $\xi$ and $\psi$.  We then used these models to compute
predicted pulse profiles in the three bands as a function of
($\xi$,$\psi$), to compare with the observed profiles.  Clearly, only
certain geometries will produce a phase shift; for example, there will
be no shift if the spot axis is nearly coaligned with both the viewing
direction and spin axis, as only one spot remains in view as the star
rotates.  Finally, we considered the effect of radiative beaming, and
repeated our analysis for a range of NS radii. In the following we
describe our procedure and results in detail.

\begin{figure*}
\hfill
\includegraphics[angle=270,width=0.47\linewidth,clip=]{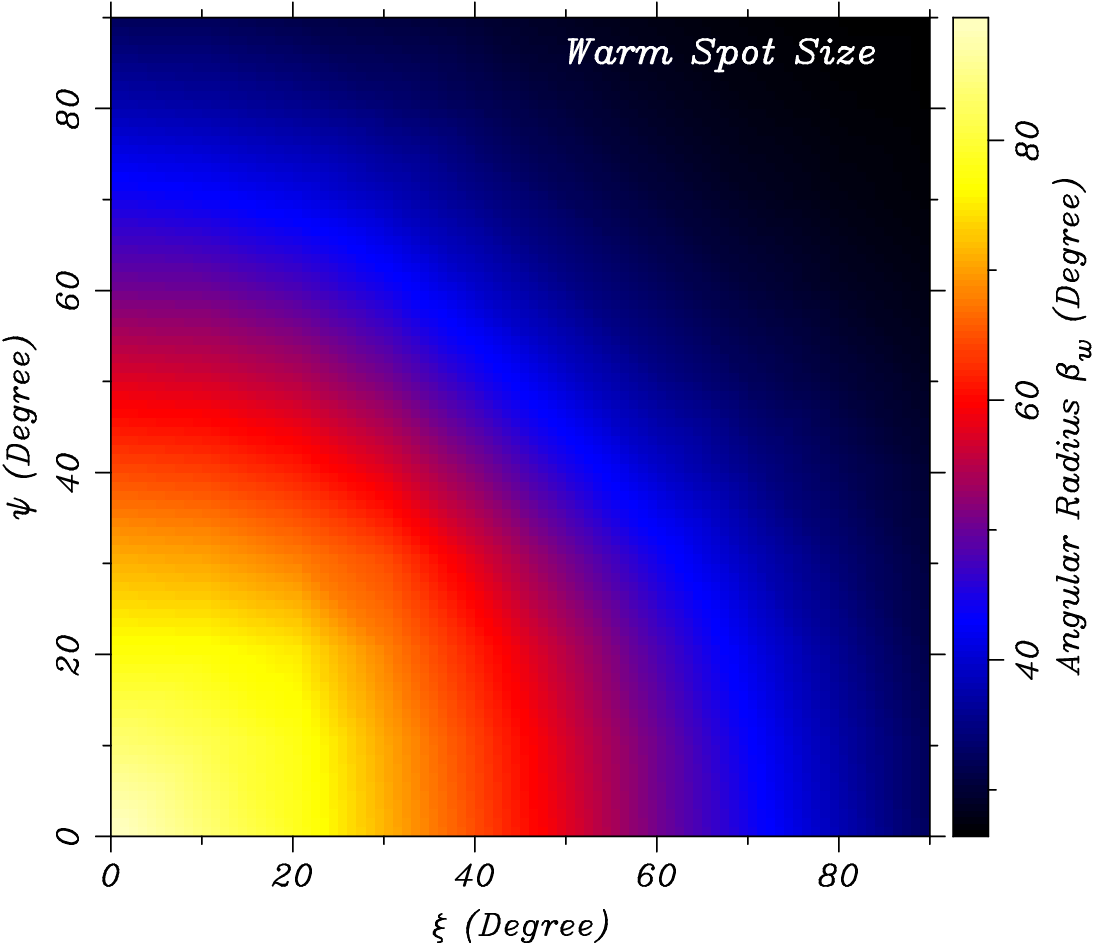}
\hfill
\includegraphics[angle=270,width=0.47\linewidth,clip=]{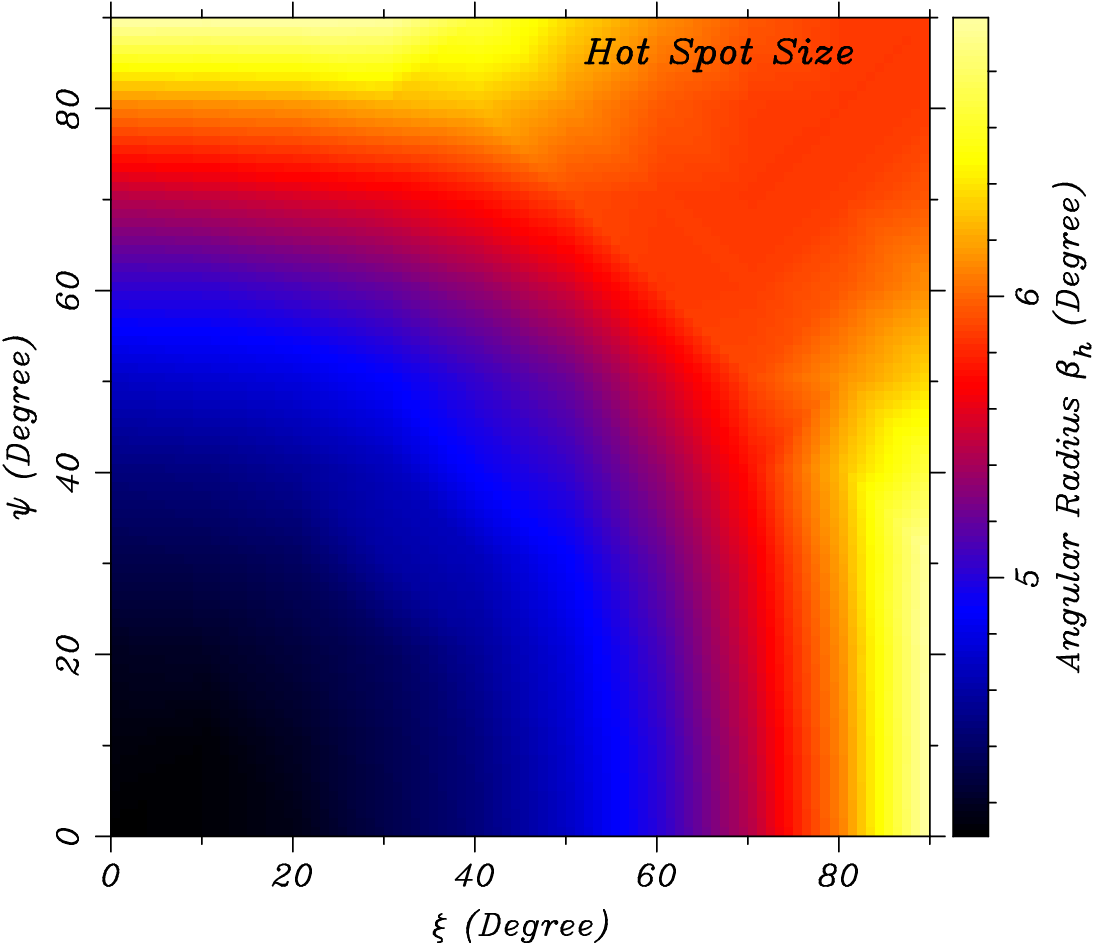}
\hfill
\caption{Best fitted values for the sizes of the two emitting spots
on the surface of \psr\ as a function of geometry
parameters ($\xi,\psi$), for a NS radius of $R=12$~km. 
Left: Map of warm-spot size parameterized by its angular radius $\beta_w$.
Right: Corresponding map for hot-spot angular radius $\beta_h$. 
The spot size is computed at intervals of $10^{\circ}$
in $\psi$ and $\xi$, and interpolated to $1^{\circ}$.
Note the very different size ranges of each spot.
Using these results, the modulation as a 
function of viewing geometry is then computed and
compared to the data, constraining the allowed geometry
as shown in Figure~\ref{fig:pf}.
This procedure is repeated for a range of NS radii,
each requiring a new pair of $\beta$ maps, with results
given in Table~\ref{tab:spectable}.}
\vspace{0.1in}
\label{fig:size}
\end{figure*}

\begin{figure*}
\hfill
\includegraphics[angle=270,width=0.47\linewidth,clip=]{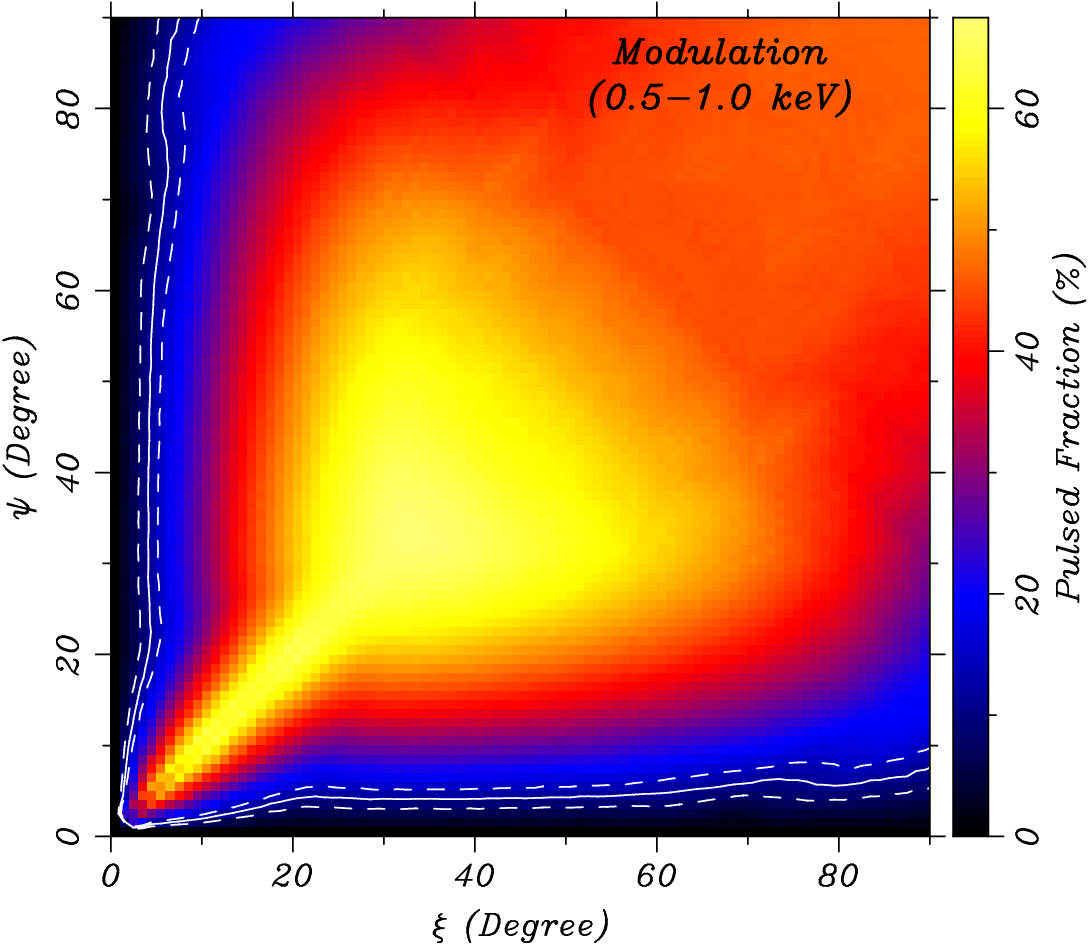}
\hfill
\includegraphics[angle=270,width=0.47\linewidth,clip=]{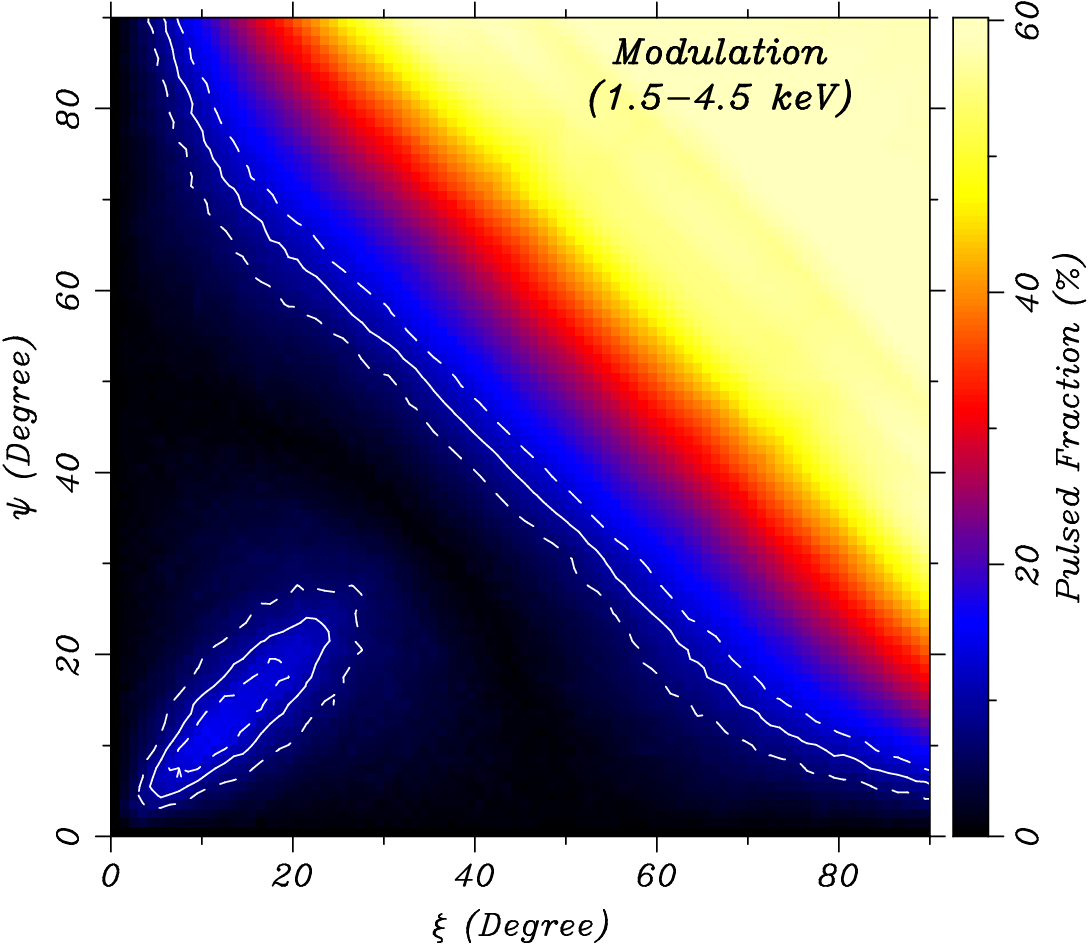}
\hfill
\caption{The model modulation derived in two energy bands 
as a function of geometry parameters ($\xi,\psi$), for a NS radius of
$R=12$~km. The pulsed fraction ranges from 0--65\% and is scaled
linearly, with yellow denoting the largest modulation.  The solid line
in each panel indicates the contour of the measured pulsed fraction in
the soft energy band (left panel) and hard energy band (right panel),
while the dashed lines give the $1\sigma$ error range.  The possible
geometry is then strongly constrained by the intersection of the two
contour regions, as shown in Figure~\ref{fig:contour}. Note that {\it
both} antipodal spots contribution to the modulation shown in each of
these plots. }
%\vspace{0.1in}
\label{fig:pf}
\end{figure*}

\begin{figure}
\centerline{
\includegraphics[angle=270,width=0.90\linewidth,clip=]{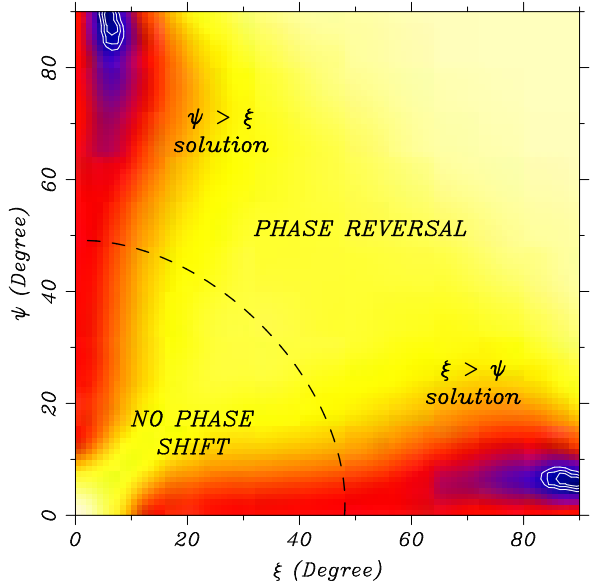}
}
\caption{
Contours of $\chi_{\nu}^2$ obtained by comparing the pulse profiles of
the antipodal model with the data in three energy bands, for a range
of angles $\xi$ and $\psi$, as described in the text.  The $1\sigma$,
$2\sigma$, and $3\sigma$ confidence levels are shown for the best
match for a NS radius of $R=12$~km.  The results are degenerate with
respect to an interchange of $\xi$ and $\psi$. The minimum
$\chi^2_\nu$ for the viewing geometry parameters is obtained at
$\psi=\bestpsi$ and $\xi=\bestxi$, evidently providing a strong
constraint. The geometries that manifest a phase shift (or not) are
indicated.}
\vspace{0.1in}
\label{fig:contour}
\end{figure}

\begin{figure}
\includegraphics[angle=270,width=0.97\linewidth,clip=]{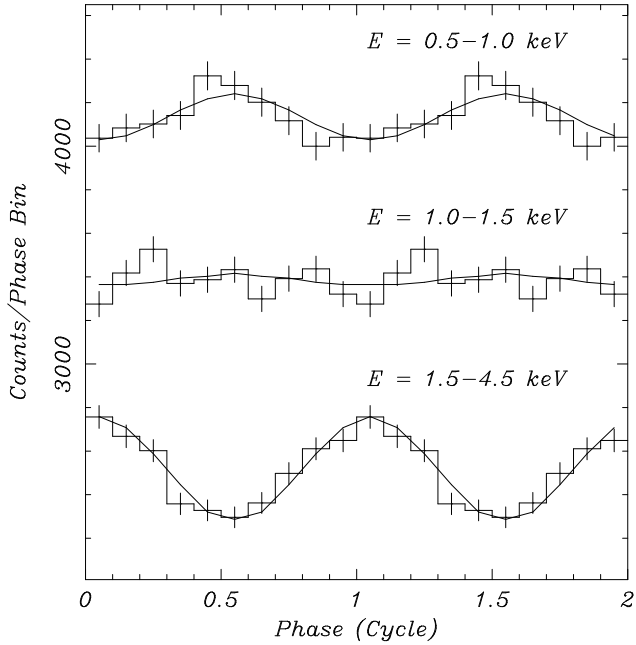}
\caption{Pulse profiles generated using the antipodal model
(solid line) that best matches the observed data (histogram)
across the three selected energy bands for \psr.  The parameters
of this model assuming a NS radius $R=12$~km are given in
Table~\ref{tab:spectable}.  The measured background in each band
has been added to the model to enable a direct comparison.}
\label{fig:profile}
\end{figure}

For a given set of viewing angles the antipodal flux model can be integrated
over phase to provide a direct comparison with the observed spectra. We have
incorporated equation~(\ref{eq:flux}) into an ``additive model'' for use in
the {\tt XSPEC} spectral fitting software \citep{arn96}.
%As a practical matter,
%we use Gauss-Legrendre weights to integrate between the (hot-spot)
%temperature boundaries modified by relativistic light bending. 
The coded model comprises 13 parameters: the NS radius and distance
($R,D$), three blackbody temperatures ($T_w,T_h,T_{\rm NS}$), two spot 
angular sizes ($\beta_w,\beta_h$), two geometrical angles ($\xi,\psi$),
the rotation phase ($\gamma$), and the Gaussian emission-line
center, width, and flux.  The column density is fixed at the best
value determined in Paper~I from a fit to the overall spectrum,
$N_{\rm H} = 4.8 \times 10^{21}$~cm$^{-2}$.
In the spectral fits, the normalization is set to unity
so that the flux is determined by $R$ and $D$, and
implicitly takes into account all relativistic effects
noted in Section~2.

\begin{figure}
\includegraphics[angle=270,width=0.97\linewidth,clip=]{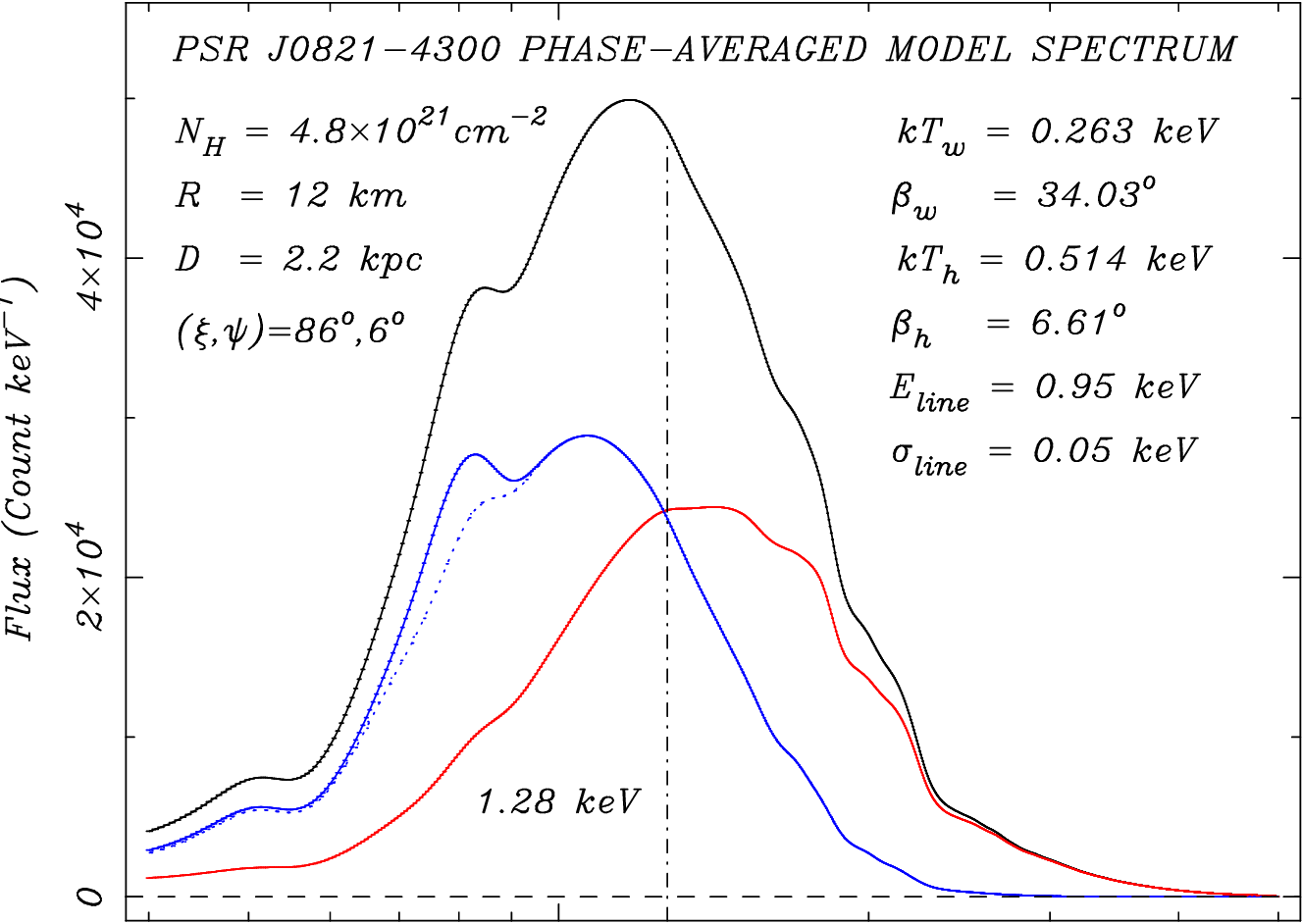}
\includegraphics[angle=270,width=0.97\linewidth,clip=]{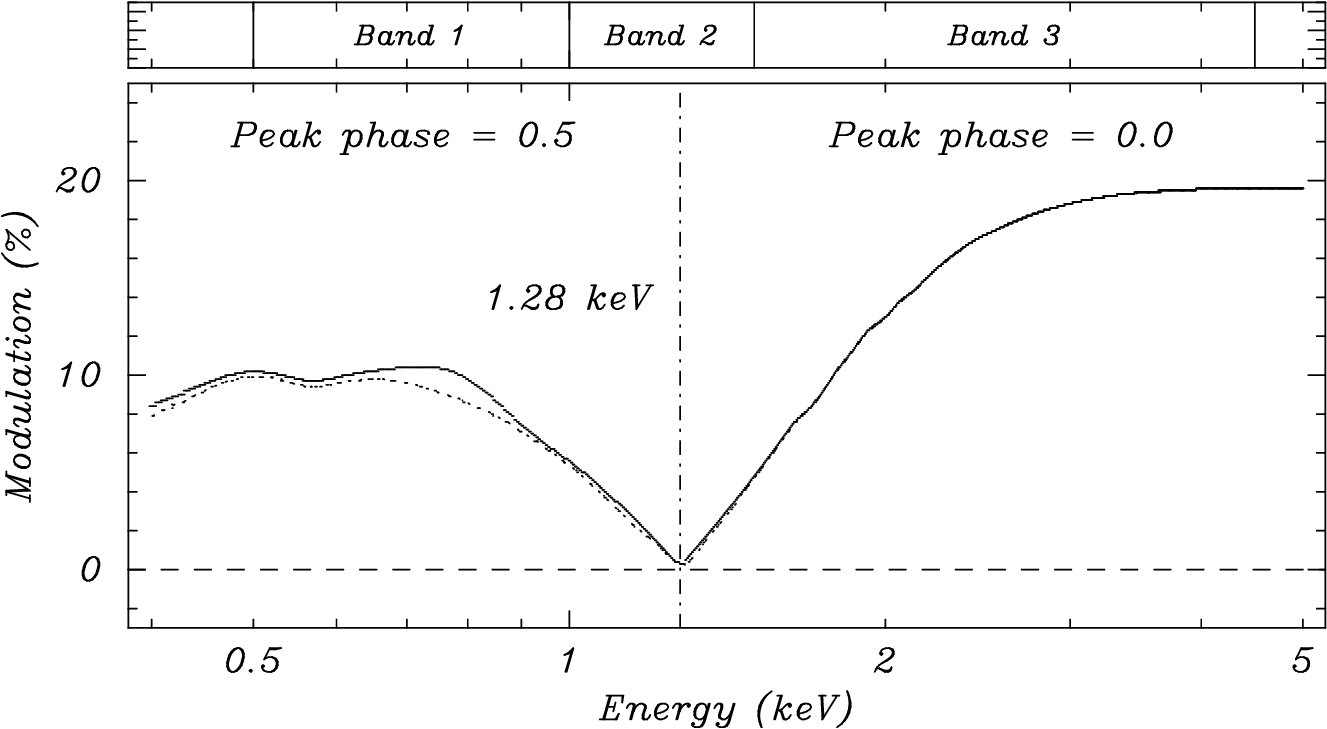}
\end{figure}

\begin{figure}
\caption{
The antipodal explanation for the observed energy-dependent
modulation and phase shift between the soft and hard X-ray bands
seen in \psr.  Top: Contribution of each phase-averaged
model spectral component to the two-blackbody model, plus a Gaussian
emission line. The dotted line shows the warm blackbody component
without the emission line contribution.  Bottom: The model pulse modulation
as a function of energy.  The strong energy dependence
results from the relative contributions of (out of phase)
 overlapping (in energy) flux from the two spectral components. The maximum modulation is
19.6\% at the highest energy.  However, the modulation in Band~3 is
smaller, as it is weighted by lower-energy photons.
The emission-line contribution is evident as the difference
between the solid and dotted lines.
The phase of the peak of the light curve follows the dominant
spectral component at each energy, and is restricted by
the symmetry of the model
to either $0$ or $0.5$ cycles.  Where the spectral components cross,
the phase must shift by $180^{\circ}$ as observed (see Figure~\ref{fig:profile} and Paper~I).}
\label{fig:explain}
\end{figure}

The model allows for a uniform temperature $T_{\rm NS}$ in the
inter-spot area, but since none is necessary for an acceptable fit to
the spectrum or pulse profiles, we set $T_{\rm NS}=0$.  Nevertheless,
to place a model-independent upper limit on $T_{\rm NS}$, we have
simulated blackbody spectra in {\tt XSPEC} for $R=12$~km and
$D=2.2$~kpc (Reynoso \etal\ 1995), increasing the temperature until
the model exceeds the spectrum of \psr\ at low energies.  Since the
model counts depend on the interstellar column density, we assumed
here the largest value measured by \citet{kas10} for filaments in the
\pupa\ remnant in {\em Chandra} data, $N_{\rm H}=5.5 \times
10^{21}\;{\rm cm}^{-2}$.  This yields a conservative $3\sigma$ upper
limit of $T^{\infty}_{\rm NS}< 0.15$~keV;
\citet{hwa08} report a significantly lower value of 
$N_{\rm H}=3 \times 10^{21}\;{\rm cm}^{-2}$ from \suzaku\
measurements, which would allow a smaller limit on $T^{\infty}_{\rm
NS}$, if applicable.

To map the antipodal model for \psr\ as a function of ($\xi,\psi$) we
generated best fit model parameters over the grid spanning
($0^{\circ}<\xi <90^{\circ};\; 0^{\circ}<\psi<90^{\circ}$) in steps of
$10^{\circ}$ by systematically fitting for $T_w$, $T_h$, $\beta_w$,
and $\beta_h$.  Since we fitted the phase-averaged spectrum (presented
in Paper~I and described therein), a reasonable approximation for the
equivalent phase-averaged model flux is $F_{\rm avg} =
[F(\gamma=0^{\circ})+ F(\gamma=180^{\circ}) ] /2$.  The flux of each
blackbody component depends on the size of its respective emission
spot as parameterized by its $\beta$, which sets the normalization for
that component.  As we consider the range of viewing orientations, the
size of the spots must be adjusted to keep the model flux fixed to
that of the observed value. This is because the projected flux from
the phase integrated emission strongly depends on the viewing geometry
($\xi,\psi$), which is not known {\it a priori}. In comparing with 
data, for each trial ($\xi,\psi$) pair, we fit for the spot sizes
($\beta_w,\beta_h$) that correctly normalize the blackbody components
to match the observed spectrum.

The grid of best fitted models with values of
($\beta_w,\beta_h,T_w,T_h$) describes the flux from \psr\ as a function
of energy and phase ($F[E_{\infty},\gamma]$)
according to equation~(\ref{eq:flux}) for any realizable
geometry\footnote{Not all values of ($\xi,\psi$)
can fit the spectrum, as the available flux in some cases,
e.g., $\xi,\psi$ both near zero, are
insufficient to match the data even with $\beta_w$ at its maximum
value of $90^{\circ}$. Such cases produce unacceptable $\chi^2$
statistics.},
the latter parameterized by ($\xi,\psi$) and a set of fixed $R,D$,
and emission-line parameters.  We find that the spot temperatures
remain almost constant in ($\xi,\psi$) and use the average value given in
Table~1 in our final fits.  The resulting fitted values of
$\beta_w$ and $\beta_h$ for the case $R=12$~km
are interpolated to $1^{\circ} \times 1^{\circ}$ pixels on a
$90^{\circ}\times90^{\circ}$ grid and displayed in Figure~\ref{fig:size}.
%In our final fits, the temperatures are held at  of $kT^{\infty}_w=0.213$~keV,
%$kT^{\infty}_h=0.416$~keV (renormalized to $R_{\infty}$).

We are now prepared to compute the set of phase-resolved modulations
in the three interesting energy bands, which manifest the phase shift.
Specifically, the pulse in the soft $0.5-1.0$~keV band peaks at
rotation phase $\gamma=180^{\circ}$, the hard $1.5-4.5$~keV band peaks
at $\gamma=0^{\circ}$, while the modulation in the intermediate
$1.0-1.5$~keV band cancels out to $< 1\%$.
The comparison is best done in counts space rather than flux,
to make use of the counting statistics for errors.  The background
counts are measured for each band and added to the model counts. 
For each ($\xi,\psi$) pair, the antipodal flux model, including
interstellar absorption, is folded through the \xmm\ EPIC response
function used for the spectral fits (see Paper~I).
For each energy band, we record the magnitude of the model
modulation $f_p$ defined as
\begin{equation}
f_p =\frac{F_{\rm tot}-N \ {F_{\rm min}}}{F_{\rm tot}}\;,
\label{eq:pf}
\end{equation}
where the flux $F_{\rm tot}$ is the total flux in the band,
$N$ is the number of bins in the light curve (ten in this case),
and $N \ {F_{\rm min}}$ is the ``unpulsed'' flux in the band,
determined from the bin of minimum flux. The results are
shown in Figure~\ref{fig:pf} for the soft and hard energy bands.

For a given ($\xi,\psi$) pair, the model profile in ten
phase bins is compared to the observed counts in the three bands,
for a total of 30 bins, using the $\chi^2$ statistic.
Since the total count normalizations in each band were fixed earlier
by the spectral fits via the $\beta$ parameters, the mean counts of
the models match the means of the three observed light curves.  This
reduces the number of degrees of freedom by three, from 30 to 27.
The ($\xi,\psi$) region that best matches the measured pulsed fraction
in the soft and hard bands, $f_p=11\% \pm 2\%$ corrected for
background, is indicated by the contours in Figure~\ref{fig:pf}.
While there is a range of allowed angles in each band separately,
they overlap in only two small regions corresponding to the minimum
$\chi^2$.  This uniquely identifies the geometry of the system, within
statistical uncertainty.
More specifically, we find that, for
$R=12$~km, the most likely geometry is specified by the combinations
$(\psi,\xi)=(\bestpsi,\bestxi)$ or $(\bestxi,\bestpsi)$.  Since
the flux depends on the angles $\xi$ and $\psi$ only through the
parameter $\alpha_h$ in equation~(\ref{eq:alpha}), it is
symmetric with respect to an exchange of $\xi$ and $\psi$, yielding
the two possible solutions. This is evident in the $\chi^2$ map that
compares the model and observed profiles in the three bands
(Figure~\ref{fig:contour}).  In one solution the spin axis is nearly
perpendicular to the line of sight, while in the other solution the
two lines are nearly parallel.  Figure~\ref{fig:profile} shows the fit
to the pulse profiles for these equivalent best solutions.
Finally, the 
parameters for the model that resulted in the smallest $\chi^2$ for
each test radius $R$ are given in Table~\ref{tab:spectable}.  In all
cases, the reduced $\chi^2_{\nu}$ is near unity, indicating an
excellent match to the observed pulse profiles.

Figure~\ref{fig:explain} provides a graphic explanation of the origin
of the observed energy-dependent modulation and phase-shift seen from
\psr.  For the set of model parameters that best fit the data,
we graph separately the phase-averaged fluxes for the two spots to
gauge their contribution to the light curve.  At the lower energies,
the large warm spot dominates the spectrum and the light curve peaks
when this component is in view, while at the higher energies, the
small hot spot dominates and it peaks in view 0.5 cycles later.  Thus,
the energy at which the dominant spectral component switches is around
1.3~keV, and the peak phase necessarily shifts by $180^{\circ}$ at
this energy due to the antipodal symmetry in longitude. This phase
reversal, at this energy, is thus a direct consequence of the crossing
of the spectral components of different temperatures. The agreement
between the energy of phase reversal and the energy of the spectral
cross-over point provides direct evidence of the correctness of the
model.

Figure~\ref{fig:explain} also shows that the predicted modulation is
19.6\% at the highest energies, where 100\% of the flux comes from the
$T_h$ blackbody component. At lower energies, the modulation is
reduced exactly in proportion to the increased contribution of the
$T_w$ or $T_h$ spot, depending on which one dominates the flux at a
given energy, $f_p(E) = 19.6 \% \times |F(E;T_w) - F(E;T_h)| /
[F(E;T_w) + F(E;T_h)]$. For our broad energy bands, used to compare
the model with data, the observed modulation is evidently weighted by
the total flux over the band. This is especially clear in the highest
energy band (1.5--4.5 keV), in which most of the photons are from the
lower end of the band where the modulation is significantly less than
19.6\%. The modulation in the middle band, which spans the phase shift
with equal counts, is mostly canceled out.  That a similar pulsed
fraction is measured in the lower and higher bands is largely
coincidence in this case. Figure~\ref{fig:explain} also shows the
contribution of the spectral line at $0.77$~keV; a small but
significant increase in the modulation results.  The photon statistics
of the current data do not allow a more detailed comparison of the
modulation as a function of energy.

Next we explored the dependence of the fitted parameters on the radius
$R$ of the NS. As discussed above, due to general-relativistic effects
$R$ is not simply a normalization of the flux; this requires us to
compute $\beta$ maps for each test value of $R$.  Therefore, we
repeated our full analysis for the range of values $9\le R \le 14$~km,
in 1~km increments.  These results are presented in
Table~\ref{tab:spectable}.  In principle, the relativistic effects can
lead to a preferred $R$, but statistically, no unique radius is
suggested here.  The general trend is that the angles $\beta_w,
\beta_h$, and $\xi$ increase with decreasing radius. Two counteracting
effects, both due to flux conservation, influence the spot sizes.
Gravitational redshift decreases the inferred emission area on a more
relativistic (smaller) star to compensate for the larger needed
surface temperature.  On the other hand, for a fixed distance $D$, the
spot angular size increases on the smaller star. For the values of the
fitted parameters here, the latter effect tends to dominate over the
former, reducing the modulation from a smaller star for the same
viewing geometry. However, the most important effect influencing the
amplitude of the modulation is the gravitational deflection of light
rays, which acts to suppress the pulsed flux for the smaller star.
Therefore, in order to reproduce the same observed level of
modulation, a smaller $R$ requires a larger modulation of the viewing
angle $\alpha(t)$, which, according to equation~(\ref{eq:alpha}), is
obtained by increasing either $\xi$ or $\psi$. This explains the
trends in Table~\ref{tab:spectable}\footnote{The larger of the two
angles ($\xi,\psi$) is unchanged as a function of radius relative to
its $2^{\circ}$ error, while the smaller one is
clearly decreasing with radius as compared to its $1^{\circ}$ error.}
%The error on $\psi$ is about $2^\circ$, while the error on $\xi$ is 
%$$1^\circ$, so the trend is towards a decrease of $\xi$.}.
%{(\bf NOTE: need to CHECK more here since there are TWO spots, and while 
%one alpha decreases, the other increases)}

We then repeated our analysis for a locally anisotropic intensity
pattern, $I(\delta) \propto \cos\,\delta$. For the nominal radius
$R=12$~km, the combined constraints from the pulsed fraction and the
phase shifts moves the best fitted viewing angles to
$(\xi,\psi)=(84^\circ,3^\circ)$, with a larger uncertainty than found
for the unbeamed case. This behavior
is readily accounted for.  Forward beaming enhances the emission
in the direction of the spot axis, hence increasing the differences in
observed flux as the axis of the spot moves toward and away from the
observer.  As a result, a fit to the observed level of modulation
requires smaller values of the viewing angles. The pulse profile is
found to remain sinusoidal and is statistically indistinguishable from
the isotropic intensity case.  For the specific geometry of \psr, such
beaming is a weak effect in narrowing the pulse profile
because of our relatively unmodulated views of the two antipodal spots
at glancing angles.

In the antipodal model, the energy dependent phase reversal is a
direct consequence of switch in dominance between the two blackbody
spectral components of different temperatures. As mentioned previously,
this agreement between the cross-over energy in phase and spectra is
distinct feature of this symmetric model, where the emission spots are
exactly antipodal.  However, similar light curves can be obtained if
the spots remain opposite in longitude, but are allowed to move closer
in latitude.  Such geometry can still produces a $180^{\circ}$ phase
reversal.  Allowing this additional degree of freedom, the angles
$(\xi,\psi$) would not be so strongly constrained as in the antipodal
model.  More generally, if the spot locations differ in longitude by
$\Delta\gamma<180^{\circ}$, such an ``offset'' model allows the
possibility of a continuous phase shift as a function of energy.  The
effect of asymmetric spot locations is clear in the energy-dependent
model profiles, but any such effect in \psr\ is not apparent, and
would require higher quality data to discern. Description of the
geometry of offset models can be found in Bogdanov \etal\ (2007, 2008),
for example.

%\clearpage
\begin{deluxetable}{lccccccc}
\tablecolumns{8}
\tighten
\tablewidth{0.0pt}
%\rotate
%\normalsize
%\large
%\small
\tablecaption{Model Results as a Function of NS Radius}
%\footnotesize
%\normalsize
\tablehead{
\colhead{Parameter} & \colhead{Unc.$^a$} & \colhead{}  & \colhead{}   & \colhead{$R$ (km)} & \colhead{}   & \colhead{} & \colhead{} \\
\colhead{}          & \colhead{}     & \colhead{9} & \colhead{10} & \colhead{11}       & \colhead{12} & \colhead{13} & \colhead{14}
}
\startdata
%\noalign{}
%{\smallskip}
$kT_w$ (keV)    & 3\%  & 0.29 & 0.28 & 0.27 & 0.26 & 0.26 & 0.25  \\ 
$kT_h$ (keV)    & 3\%  & 0.57 & 0.54 & 0.53 & 0.51 & 0.50 & 0.50  \\ 
$\beta_w$       & 6\%  & $39\!^{\circ}$ & $37\!^{\circ}$ & $35\!^{\circ}$ & $34\!^{\circ}$ & $32\!^{\circ}$ & $31\!^{\circ}$ \\ 
$\beta_h$       & 11\% & $ 8\!^{\circ}$ & $ 7\!^{\circ}$ & $ 7\!^{\circ}$ & $ 7\!^{\circ}$ & $ 6\!^{\circ}$ & $ 6\!^{\circ}$ \\ 
$\xi$ or $\psi$ & $2^{\circ}$  & $86^{\circ}$ & $87^{\circ}$ & $87^{\circ}$ & $86^{\circ}$ & $87^{\circ}$ & $87^{\circ}$   \\ 
$\psi$ or $\xi$ & $1^{\circ}$ & $ 9^{\circ}$ & $ 7^{\circ}$ & $ 7^{\circ}$ & $ 6^{\circ}$ & $ 5^{\circ}$ & $ 5^{\circ}$   \\ 
$\chi^2_{\nu}$~(27 DoF) &\dots & 0.94 & 1.00 & 0.99 & 1.00 & 0.95 & 0.97
\enddata
\tablecomments{Spectral fits with fixed parameters
$N_{\rm{H}}=4.8\times 10^{21}$~cm$^{-2}$ and $D=2.2$~kpc.  Gaussian
line model parameters are fixed at energy $E^{\infty} = 0.77$~keV,
width $\sigma=0.05$~keV, and flux normalization
$1.8\times10^{-4}$~ph~cm$^{-2}$~s$^{-1}$.
\\$^a$ The $1\sigma$ uncertainties in the spectral parameters are estimated by running the
XSPEC {\tt error} command; the uncertainties in $\xi$ and $\psi$ are
determined from the $\chi^2$ map of Figure~\ref{fig:contour}.}
%\vspace{-0.1in}
\label{tab:spectable}
\end{deluxetable}

%\clearpage

\section{Discussion of Model Results}

Using an antipodal spot model, we have accounted for all of the
details of the \xmm\ observations of \psr\ described in
Paper~I.  In particular, we can reproduce the overall spectral shape,
energy-dependent pulsed modulation, and abrupt $180^{\circ}$ phase
reversal at the cross-over energy of the fitted blackbody components. 
In so far as no observed phenomena remains unmodeled, and
no unobserved features are predicted, the antipodal model provides a
credible description of the geometry of emission from
the CCO in \pupa.
The full data set can be reproduced, with slight differences
in the best fitted parameters, assuming either isotropic or
forward-beamed emission.  Differentiating between these assumptions
will require observations with higher statistics.

By matching the observed modulation in three broad energy bands, we
are able to restrict the angles that the hot-spot axis and the line of
sight make with respect to the spin-axis to within $<2^{\circ}$, up to
the degeneracy between these two angles.  Either the spin axis lies
nearly parallel ($\bestxi$) to the line-of-sight, with the hot-spot
axis at $\bestpsi$, or the hot-spot axis is nearly co-aligned with the
spin-axis, but perpendicular to the line-of-sight.  In the absence of
a strong physical motivation to prefer one of these configurations
over the other, we note that the a priori probability of the spin axis
lying $\bestxi\pm 1^{\circ}$ from the line of sight is only $3.6
\times 10^{-3}$, while the probability that it is at $\bestpsi\pm
2^{\circ}$ is $7.0 \times 10^{-2}$, a factor of 20 larger, although
still small.  We note that the specific orientations that fit the
observations of \psr\ are not the only ones that allow phase reversals
in the two-temperature model.  Rather, phase reversals are found in
the majority of configurations of Figure~\ref{fig:contour}.

Comparing with the other CCO pulsars, we see that PSR J1852+0040
in the SNR Kesteven 79 \citep{hal10} also has a two-temperature
X-ray spectrum, but its highly modulated pulse ($f_p = 64\%$)
is single-peaked and virtually invariant with energy.
Because of this, its emitting regions are likely
to be concentric, or nearly so.  In the case of 1E~1207.4$-$5209
in \pkssrc, there are large variations in pulse phase
and amplitude as a function of energy \citep{pav02b,del04},
with the largest pulsed fraction coinciding with the strong 
absorption lines in the unique spectrum of this pulsar.
This effect may be a manifestation of
angle-dependent scattering in cyclotron lines, which is 
the favored identification of the spectral features considering
the upper limit of $B_s < 3.3 \times 10^{11}$~G on the surface
dipole field from the absence of spin-down \citep{got07}.
The data on 1E~1207.4$-$5209 should be fitted
with detailed atmosphere models that include quantum treatment
of the cyclotron harmonics \citep{sul10}.

Applying this model to surface thermal emission from CCOs with
weak magnetic fields (anti-magnetars) is an especially apt use,
in that additional complicating emission mechanisms that are evident
in other classes of pulsars (see Section~5)
are absent in CCOs.  Such contributions
include nonthermal magnetospheric emission in spin-powered pulsars,
polar-cap heating from backflowing particles, and transient and
variable heating from magnetic field decay in magnetars.
The first two extra contributions can be
significant even for recycled millisecond pulsars,
which are now known to be efficient $\gamma$-ray emitters
\citep{abd09a,abd09b}.
Observations and upper limits on spin-down of CCOs indicate
spin-down luminosities that are smaller than their thermal
X-ray luminosities, and dipole magnetic fields of order $10^{10-11}$~G,
remarkably small for young pulsars.  These properties imply that
none of the above-mentioned emission and surface heating mechanisms
can be significant, and 
constrain the effects that may be responsible for the
multiple temperatures that are a ubiquitous feature of CCO spectra,
even those that have not yet been observed to pulse.

For the assumed distance to \psr\ of 2.2~kpc and a radius of 12~km,
the best match for the modulation fixes the extent of the hot and warm
regions to angles $\beta_h = 6.\!^{\circ}6\pm0.\!^{\circ}5$ and
$\beta_w = 34.\!^{\circ}0\pm2\!^{\circ}$, representing $0.33\%$ and
$8.5\%$ of the surface area, respectively.  The existence of a hot
spot is difficult to understand in the context of a weakly magnetized
NS, as it requires a mechanism to confine the heat to such a small
area.  Using \xmm, we obtained a new period measurement of \psr\ on
2010 May~2 using the identical observational setup and analysis as
described in Paper~I. We obtained another measurement of the
pulsations on 2010 Aug~16, from a \chandra\ CC-mode observation.
These results will be presented in a future publication.  The period
is found to be unchanged from the values observed in 2001.  In
combination with the previous measurements listed in Paper~I, this
places a $2\sigma$ limit on $\dot P$ of $<3.5 \times 10^{-16}$
and, under the assumption of dipole spin-down,
$B_s < 2.0 \times 10^{11}$~G, confirming \psr\ as an
anti-magnetar.  Given the corresponding upper limit on spin-down
luminosity of $<1 \times 10^{34}$ erg~s$^{-1}$, the hot-spot
luminosity of $\approx 2 \times 10^{33}$ erg~s$^{-1}$ can hardly be
attributed to external heating by backflowing particles.  The same
problem was discussed in the context of the highly pulsed emission
from PSR J1852+0040 in Kes~79 \citep{hal10}.

Possible explanations for the properties of CCOs are largely
focused on magnetic field induced anisotropies
in the surface temperature of a NS, as proposed by \citet{gre83},
in which strongly enhanced conductivity in the direction parallel 
to the magnetic field is matched by a corresponding reduction in the
perpendicular direction. 
The effect of the magnetic field on the heat transport of the
crust and envelope of neutron stars has been investigated by a number
of authors \citep[e.g.,][]{hey98,hey01,pot01,lai01,gep04,gep06,per06a,pon09}.
While heat transport in the core ($\rho\ga
1.6\times 10^{14}$ g~cm$^{-3}$) is expected to be roughly isotropic
due to proton superconductivity, anisotropy of heat
transport becomes pronounced in the outer envelope
($\rho\la 10^{10}$ g~cm$^{-3}$) for field strengths $B\ga 10^{10}$~G, 
and it extends deeper into the whole crust for higher fields,
$B\ga 10^{12}-10^{13}$~G.  The main question is whether subsurface
fields in CCOs can be strong enough to affect heat transport to
the extent required, while not exceeding the weak external dipole
field as constrained by their spin-down properties.

\citet{gep04} discussed the differing effects of a poloidal magnetic
field in the core of the NS, versus one confined to the crust,
the true configuration being a matter of uncertainty.  From a core field,
any surface temperature anisotropy is expected to be small, while
a tangential crustal field insulates the magnetic equator and conducts
heat to the magnetic poles.  A tangential crustal field may be indicated
for CCOs, because it can lead to small hot regions where
the field emerges normal to the surface, while contributing
very little to the external dipole field.  Of particular
interest here, \citet{gep06} found that, if the crustal field
consists of both a dipolar poloidal and a toroidal component,
then configurations can be realized in which two warm regions of
{\em different sizes} are separated by a cold equatorial
belt.  However, their case study included large poloidal
magnetic fields, $B \ga 10^{12}$~G in both core and crust components,
which would tend to violate the observed
spin-down limit of $B_s < 2.0 \times 10^{11}$~G for \psr.

\section{Comparison with Other Pulsars}

The ultimate goal of this field is to infer the equation of state and
measure the radius of the NS.
%Ultimately, modeling such as this should infer the equation of state and 
%measure the radius of the NS, as well as its surface composition.  
Some progress on these fronts has been made with high-quality data
from millisecond pulsars (MSPs).  Using an unmagnetized hydrogen
atmosphere model fitted to the spectra and pulse profile of the
nearest known MSP J0437$-$4715, \citet{bog07} derived $6.8 < R <
13.8$~km (for $M = 1.4\;M_{\odot}$). \citet{bog08,bog09} also obtained
lower limits on $R$ modeling X-ray observations of MSPs J2124$-$3358
and J0030+0451.  Blackbody emission was not able to fit the pulse
profiles, thus requiring a NS atmosphere.
\citet{bog07} assumed an identical pair of
polar caps, but fitted two temperatures to each,
as required by the data, which can be understood as non-uniform
heating by backflowing particles from the magnetosphere
giving a concentric temperature gradient, as originally
modeled by \citet{zav98}. \citet{bog07}
concluded that the magnetic dipole is not centered on the star,
but must be offset by $\sim 1$~km to account for
an asymmetry in the observed pulse profile.  The data on
 \psr\ in \pupa\ are not yet of a quality to search for such effects.
On the other hand, the geometrical angles ($\xi,\psi$)
are not nearly as well constrained in the MSPs. 
(In the case of the binary MSP J0437$-$4715.
it could be assumed that $\psi=42^{\circ}$ because that
is the inclination angle of its binary orbit.)

Pulsed light curves of the middle-aged pulsars PSR B0656+14, B1055$-$52, and
Geminga, whose X-ray spectra are dominated by surface thermal emission,
have been modeled by \citet{pag95a}, \citet{pag95b}, \citet{pag96},
and \citet{per01}.  Beginning with \rosat\ data,
it appeared that PSR B0656+14 \citep{pos96,gre96}, PSR B1055$-$52 \citep{ogl93},
and Geminga \citep{hal93} had two thermal components with pulse-phase shifts
of between 0.1 and 0.3 cycles, the hotter component being attributed to a
heated polar cap.  Follow-up observations at higher energy with {\it ASCA\/}
found that the harder components from Geminga \citep{hal97} and
PSR B1055$-$52 \citep{wan98}
are better fitted by non-thermal power laws.  As beamed emission
from the magnetosphere, their hard X-ray pulses need not bear a simple phase
relationship to the soft thermal components.  Only PSR B0656+14 continued
to have two clear temperatures when observed at higher energy
\citep{pav02a}, with only a weak nonthermal tail.

Detailed study of the energy-dependent pulse profiles of these
primarily thermal pulsars with \xmm\ \citep{del05} confirm that 
PSR B0656+14 has two thermal components, with the hotter one
interpreted as a small polar cap, shifted in phase by $\sim 0.2-0.3$
cycles from the softer emission.   The spectrum of PSR B1055$-$52 
was fitted with two temperatures and a non-thermal power law,
although it is difficult to explain why the hotter blackbody component
has a pulsed amplitude of $\sim 100\%$.
A case for a hot polar cap on Geminga was made by \citet{car04}
and \citet{del05}, but \citet{jac05}, analyzing the same data,
did not find it to be necessary.
Such a component does not make a significant contribution to the spectrum
of Geminga at any energy, and its fitted pulse profile appears to have
the same phase and similar shape as the power-law component,
suggesting that it is a distinction without a difference.

Despite these difficulties, the pulsed amplitudes and fitted areas
of thermal X-ray emission from cooling neutron stars indicate that
most have highly nonuniform surface temperatures that may be
regulated by their crustal magnetic field geometry.  For example,
the \xmm\ observation of the middle-aged pulsar B1706$-$44
\citep{mcg03} shows an asymmetric, double-peaked pulse profile whose
$T^{\infty} = 8 \times 10^5$~K spectrum is compatible with the full NS area,
while having a 22\% pulsed fraction.
In contrast, PSR J0538+2817 appears to have only a hot polar cap of
$T^{\infty} = 2.2 \times 10^6$~K \citep{mcg04}.  One of the most
unusual results is the apparently thermal ($T^{\infty} = 2.4 \times 10^6$~K)
spectrum of the  high $B$-field ($4.1 \times 10^{13}$~G) PSR~J1119$-$6127
\citep{gon05}, which has pulsations of amplitude $74\% \pm 14\%$ that
are only detected below 2~keV.

Another family of thermally emitting pulsars are the nearby, isolated
neutron stars (INSs) \citep{hab07,kap09} with periods of $3.4-11.4$~s,
and pulsed fractions that range from 1.2\% for RX J1856.6$-$3754
\citep{tie07} to 52\% for RX~J1308.6+2127 \citep{sch05}.  The latter
authors fitted the double-peaked pulse profile of RX~J1308.6+2127 to a
model of two small spots with temperatures of $kT_1^{\infty} = 92$~eV
and $kT_2^{\infty} = 84$~eV separated by $~\sim 160^{\circ}$ in phase.
Timing measurements have revealed that these INSs have somewhat larger
dipole magnetic fields than most young pulsars, with $B_s \ga
10^{13}$~G \citep{kap09}, and may be significantly heated by
continuing magnetic field decay \citep{pon09}.  In this sense, they
may be the $\sim 10^6$~year old descendants of magnetars.  Several of
their spectra have very broad absorption features that have been
interpreted as ion cyclotron lines or, in the case of multiple
features, possibly atomic lines (Haberl 2007; Schwope \etal\ 2007).
One of the best studied objects of this class, RX J0720.4$-$3125,
shows a pulse phase shift of $\sim 0.1$ between soft and hard X-rays
\citep{cro01}, which suggests that there could be two spots of
different temperatures.  However, this interpretation is complicated
by long-term (years) changes in the shape of the spectrum and pulse
profile \citep{dev04,hoh09}, which lends support to the idea that,
similar to the case of magnetars, localized and variable heating by
magnetic field decay is responsible for relatively short-lived surface
thermal structure.  This is evidently the case for the transient
magnetar XTE J1810$-$197, whose declining hot spot temperatures and
areas were modeled by \citet{got05,got07}, \citet{ber09},
\citet{alb10}, and by \citet{per08}, who used a similar treatment as
that presented herein.

\citet{per06b}, \citet{zan06} and \citet{zan07} investigated models for INSs
involving a combination of star-centered dipole and quadrupole magnetic field
components to explain their asymmetric pulse profiles.
The properties of \psr\ in \pupa\ may ultimately be ascribed to these same
effects, but in a simpler system that is not variable in time.

\section{Conclusions}

We modeled the \xmm\ spectra and pulsed light curve of
\psr\ in \pupa, one of three CCO pulsars whose dipole magnetic
field strengths are measured to be less than those of all spin-powered
pulsars of similar age ($B_s < 2.0 \times 10^{11}$~G in the case of
\psr).  The sizes and configurations of the surface hot and warm spots
on \psr\ are particularly well constrained.  The two emitting areas
differ by a factor of 2 in temperature and 20 in area, which
conveniently endows them with similar luminosities that fall in the
\xmm\ bandpass.  The $180^{\circ}$ phase reversal between the soft and
hard X-ray pulse profiles reveals the antipodal geometry.  It is
especially significant that the X-ray spectra and pulse profiles of
CCOs indicate considerably nonuniform surface temperatures.  Many of
the mechanisms that are held responsible for such effects in other
classes of NSs are not expected to be operating in these
anti-magnetars, which appear to be simple cooling neutron stars whose
conduction of heat from the interior is highly anisotropic.  The
essential problem in understanding CCOs is to explain how this is
accomplished without creating a strong external dipole magnetic
field.  Our tentative hypothesis is that even CCOs have strong
tangential fields buried in the crust that channel heat toward the
magnetic poles, or external quadupole fields.  But what is the geometry of that
magnetic field?  Although the orientation of the hot spots in \psr\ is
determined to within $2^{\circ}$, the degeneracy of the model does not
allow us to decide if the axis of the hot spots is nearly aligned with
the NS spin axis or nearly perpendicular to it.  Geometrical
probability, as well as the observation of larger pulsed fractions in
other NSs, would suggest the latter.  The actual geometry is probably
fixed during the genesis of magnetic fields in these young NSs, which,
in the case of CCOs, have not spun down since their birth and are
likely to have preserved the natal $B$-field configuration.

\acknowledgements

We thank the referee, Silvia Zane, for helpful comments on the paper.
This work is based on observations obtained with \xmm, an ESA science
mission with instruments and contributions directly funded by ESA
Member States and NASA.  This work was supported by NASA XMM grant
NN08AX71G.

\end{document}